# Interplane relaxation and the bilayer coupling in $Y_2Ba_4Cu_7O_{15}$

H. Monien and A. W. Sandvik *

*Theoretische Physik, ETH Zürich, CH-8093 Zürich, Switzerland*
*\* University of California Santa Barbara, Santa Barbara, CA 93106*

*The origin of the spin gap in the underdoped cuprate superconductors is still mysterious. Experimental evidence from neutron scattering and NMR experiments indicates that the spin gap might be present only in the bilayer compounds. A naive calculation for a two plane Heisenberg model locates the order-disorder transition only for very large exchange coupling between the bilayers. We propose a interplane relaxation experiment which might quantitatively estimate the strength of the bilayer coupling. We make detailed predictions for the size and the temperature dependence of the interplane relaxation rate.*

## 1. INTRODUCTION

Recently the low energy spin dynamics of the high $T_c$ superconductors has attracted a lot of interest. Some of the theories for the high temperature superconductors are based on low frequency spin fluctuations.[1] The unusually strong temperature dependence of the measured spin susceptibility, $\chi_s(T)$ of some of these compounds ($YBa_2Cu_3O_{6.63}$, $YBa_2Cu_4O_8$, ...) rules out a Fermi liquid like ground state with a temperature independent spin susceptibility. The decreasing spin susceptibility indicates the closeness of a critical point. Many people have proposed that this critical point should be identified with the order-disorder transition in the two dimensional Heisenberg Quantum Antiferromagnet.[2] The reason why these systems are close to this transition is however still under discussion. Millis and Monien[3] proposed that the bilayer coupling is driving the system towards the quantum disordered phase whereas Sokol and Pines[4] proposed that doping alone is sufficient to reach the quantum disordered phase. This critical point has recently been studied in great detail by Sachdev et al. and others.[5–8] Millis and Monien concluded from the analysis of existing experimental NMR data that the spin gap only appears in systems containing $CuO_2$ bilayers. An additional hint that bilayer correlations are important in YBCO comes from the neutron scattering experiments by Tranquada et al.[9] which show nearly complete antiferromagnetic correlations of the spins in the bilayers. All this points in the direction of a sizable exchange coupling between the planes. The study of a model of two antiferromagnetically coupled Heisenberg planes reveals that the exchange coupling between the planes, $J_\perp$ has to be of the order $2.5 \times J$,[7,8,10] the antiferromagnetic exchange coupling in the plane, which is of the order of $J \sim 1000K$. This exchange coupling seems too large for a realistic material. Several groups[11-13] have attempted to clarify mechanisms which could lead to an enhancement of the singlet correlations between the bilayers. In this paper we analyze theoretically in more detail a recently proposed experiment,[14] which makes it possible to directly measure the transition of the "correlated plane regime" to the "uncorrelated plane regime" and to obtain an estimate for the exchange coupling between the planes $J_\perp$.

## 2. INTERPLANE RELAXATION

The basic feature of all CuO superconductors are the CuO planes. In $YBa_2Cu_3O_{6+\delta}$ two layers form a so called bilayer which is separated from the next bilayer by a large distance. Each of the layers of the bilayers is completely equivalent in $YBa_2Cu_3O_{6+\delta}$. Recently a new CuO superconductor has been synthesized in which the $CuO_2$ layers of the bilayers are not equivalent. The unit cell contains two building blocks one of which has one chain, similar to $YBa_2Cu_3O_7$ and the other two chains, like $YBa_2Cu_4O_8$. The structure is sketched in Fig. 1. The

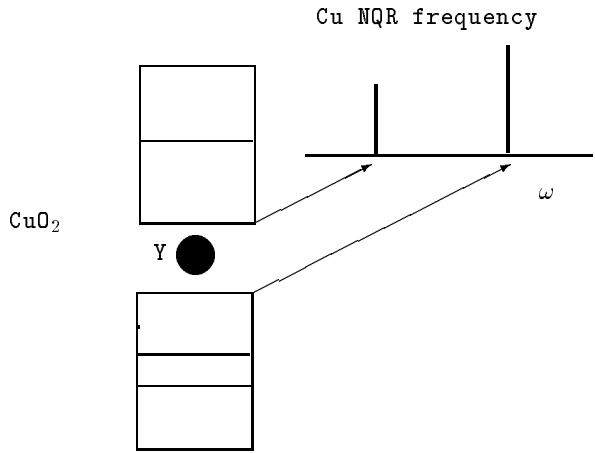

Fig. 1. Schematic structure of $Y_2Ba_4Cu_7O_{15}$

important point is that each of the $CuO_2$ layers are attached to a different chemical environment but are otherwise left intact. One plane of the bilayer is attached to the single chain block the other one to the double chain block. Some of the properties of this material have been studied by Stern et al.[15] In particular they find that the nuclear quadrupole resonance (NQR) frequencies of the Cu nuclear spins in the individual planes of the bilayers are different from each other. This allows one to access the spins in both $CuO_2$ planes seperately. It is known that the transverse relaxation rate, $1/T_2$, is dominated by an indirect exchange coupling via the conduction electrons.[16] A Cu nuclear spin at site $i$ in plane $(a)$, couples to an

electron spin in the same plane, $(a)$, via the hyperfine Hamiltonian

$$H_{hf} = A_{\alpha\alpha} I_{i\alpha}^{(a)} S_{i\alpha}^{(a)} + B \sum_j I_{i\alpha}^{(a)} S_{j\alpha}^{(a)} \qquad (1)$$

where $A_{\alpha\alpha}$, $\alpha \in x, y, z$, is the on-site hyperfine coupling for the $Cu^{2+}$ spin and $B$ is the isotropic transferred hyperfine coupling. The sum over $j$ extends over the four neighboring Cu sites in the <u>same</u> plane. The nuclear Cu spin polarizes the neighboring electronic spins in the same plane which in turn produce an additional field for the neighboring Cu nuclear spins. This process generates a nuclear-spin nuclear-spin interaction of the form

$$H_{II} = \sum_{ij} I_{i\alpha}^{(a)} V_{\alpha\alpha}(R_i - R_j) I_{j\alpha}^{(a)} \qquad (2)$$

$V_{\alpha\alpha}(R_i - R_j)$ is defined by its Fourier transform:

$$V_{\alpha\alpha}(q) = F_{\alpha\alpha}(q) F_{\alpha\alpha}^*(q) \chi_s^{(a,a)}(q) \qquad (3)$$

where $F_{\alpha\alpha}(q)$ is the Fourier transform of the hyperfine coupling constant in Eq. (1) and $\chi_s^{(a,a)}(q)$ is the static spin susceptibility for a single plane. For a $Cu^{2+}$ spin in a $d_{x^2-y^2}$ orbital the hyperfine interaction is anisotropic and largest in the c direction perpendicular to the planes. The Cu nuclear spin interaction, Eq. (2), dominates the dipol-dipol interaction of the nuclear spins because the spin suscepti­bility, $\chi_s(q)$, is strongly enhanced at the antiferromagnetic wavevector in the plane. The interaction is giving the main contribution to the dephasing time of the Cu nuclear spins in one $CuO_2$ layer. Pennington and Slichter[16] derived an expression for the dephasing time $T_2$ for the case that the hyperfine coupling in the c direction is much larger than any other direction:

$$\left(\frac{1}{T_2}\right)^2 = n_m \left\{ \sum_q \left[F_z(q)^2 \chi_q^{(a,a)}\right]^2 - \left[\sum_q F_z(q)^2 \chi_q^{(a,a)}\right]^2 \right\} \qquad (4)$$

where $n_m$ is the density of the active NMR nuclei. The second term cancels the on-site contribution of the first term. This time has been measured in $YBa_2Cu_3O_7$ and is in reasonable quantitative agreement with simple phenomenological theories.[16,17] In the special bilayer system discussed above it is possible to excite a spin in one plane and measure the response in the other plane. Then one can use Eq. (4) to define a interplane-relaxation time for the bilayer system. In Eq. (4) $\chi_q^{(a,a)}$ has to be replaced by $\chi_q^{(1,2)}$ where the spins are sitting in different planes. In this case we do not have to subtract the on-site term. To discuss the the interplane relaxation time we use the two plane Heisenberg model which exhibits a order-disorder transition as a function of the between plane coupling $J_\perp$.

Many physical properties of the two plane Heisenberg model in the disordered phase can be calculated in an approximate way with the Schwinger-boson mean-field theory. Schwinger-boson mean-field (SBMFT) does not reproduce the numerical value of the critical exchange coupling between the planes, $J_\perp$ (numerical value

$J_\perp \approx 2.5 J$[8,10,18] vs. $J_\perp \approx 4.5 J$ in the SBMFT[19]) but it does correctly give the main features of the phase diagram. The spin-correlation functions and a number of physical properties of the two plane Heisenberg model have been obtained in this approach. Here we are interested in the calculating the correlation function of a spin in one plane, $\vec{S}^{(1)}$ with a spin in the second plane, $\vec{S}^{(2)}$. The SBMFT is rotationally invariant in spin space. We therefore can confine ourselves to calculating the $\langle S^z S^z \rangle$ correlation function. An external field in the $z$ direction $S^z$ is coupling to the spin, $\vec{S}_i^{(a)}$, in plane $(a)$, site $i$ via the Hamiltonian

$$\Delta H = \sum_{ia} h_i S_i^{(a)} \tag{5}$$

The elementary excitations of the system are the optical and the acoustic spin-waves. It is convenient to rewrite the Eq.(5) in terms of these excitations:

$$\Delta H = \sum_q \left\{ \frac{1}{2} \left( h_q^{(1)} - h_q^{(2)} \right) O_q^a + \frac{1}{2} \left( h_q^{(1)} + h_q^{(2)} \right) O_q^s \right\} \tag{6}$$

where $O_q^a$ and $O_q^s$ are operators creating spin fluctuations symmetric and antisymmetric under interchange of the planes respectively. The only nonvanishing correlation functions are $< O_q^s O_{-q}^s >$ and $< O_q^a O_{-q}^a >$. In linear response theory we obtain for the spin-correlation function when both spins are sitting in the same plane:

$$\left\langle S_q^{z(1)} S_{-q}^{z(1)} \right\rangle (q,\omega) = \frac{1}{4} \left( \chi^{ss}(q,\omega) + \chi^{aa}(q,\omega) \right) \tag{7}$$

where $\chi^{aa}(q,\omega)$ and $\chi^{ss}(q,\omega)$ are defined as

$$\chi^{aa}(q,\omega) = \int_0^\infty e^{i\omega t} \left\langle [O_q^a(t), O_{-q}^a] \right\rangle dt \tag{8}$$

$$\chi^{ss}(q,\omega) = \int_0^\infty e^{i\omega t} \left\langle [O_q^s(t), O_{-q}^s] \right\rangle dt \tag{9}$$

and the index (1) is referring to the plane one. The spin-correlation function for both spins sitting in plane (2) is of course identical.

For the spin-correlation function with the two spins in different planes we obtain:

$$\left\langle S_q^{z(1)} S_{-q}^{z(2)} \right\rangle (q,\omega) = \frac{1}{4} \left( \chi^{ss}(q,\omega) - \chi^{aa}(q,\omega) \right) \tag{10}$$

Now we are in the position to compare the interplane relaxation time, $T_{2\perp}$ with the transverse spin relaxation time $T_2$. At low temperatures the planes are coupled and only the acoustic mode is populated. In this regime $\chi^{ss}$ is much larger than $\chi^{aa}$. Comparing Eqs. (7) and (10) and we see that $T_2$ and $T_{2\perp}$ are identical. With increasing temperature the optical mode is more and more populated until, at approximately $T \approx J_\perp$ the planes are decoupling and the optical and acoustic mode merge. In this temperature regime $\chi^{aa}$ and $\chi^{ss}$ are equal.

This transition is of course an artifact of the SBMFT and will be a interplaneover if fluctuations are taken into account. For this reason we have studied the two plane Heisenberg model using a recent improvement of the Hanscomb Quantum Monte Carlo method which is particularly suited for non frustrated systems.[20] We have calculated the interplane relaxation time for various values of $J_\perp$. Fig. 2 shows the interplane relaxation time as a function of temperature for a $J_\perp = 2.5J$, close to the order-disorder transition. In this calculation we have used the values for the hyperfine coupling as determined from other NMR experiments. The interplane relaxation rate is strongly decreasing with increasing temperature. There is no strong feature visible at the temperature of $T \approx 2.5\ J$ where the Schwinger-boson mean-field theory would predict the transition from the coupled plane to the decoupled plane regime. This has several reasons. The acoustic mode is becoming broad with increasing temperature and starts to overlap with the optic mode. At a temperature of 2.5 $J$ it is very natural not to see a sharp transition. One might hope that for smaller values of $J_\perp$ the crossover is more pronounced. The value of the interplane relaxation rate, $1/T_{2\perp}$, for a $J_\perp/J \approx 2.5$ is much larger than the preliminary experimental findings by Mali and Stern.[21] We are currently exploring smaller values of $J_\perp$. The correlations between the planes seem to be strong even for rather small values of $J_\perp$.

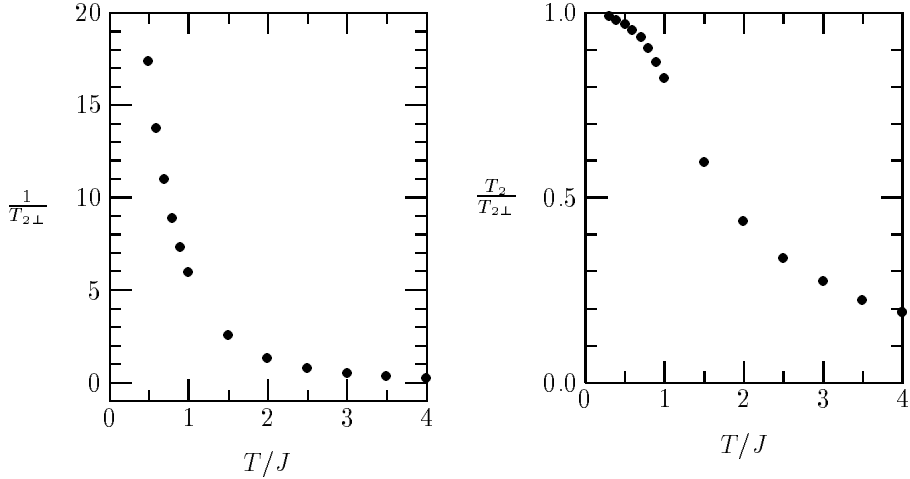

Fig. 2. Temperature dependence of the interplane relaxation rate $1/T_{2\perp}$ for $J_\perp/J = 2.5$ on a $2 \times 32 \times 32$ lattice and the ratio of of the interplane relaxation time, $T_{2\perp}$, to the transverse nuclear relaxation time, $T_2$.

## 3. CONCLUSIONS

We have analyzed an interesting experiment which might be able to pin down the value of the bilayer coupling in YBCO. The preliminary results of Mali and Stern[21] seem to indicate that $T_{2\perp}$ is of the order of $T_2$. Our recent calculation show that only a modest value of $J_\perp$ is required to obtain a large $T_{2\perp}$. The values for $J_\perp$ are in agreement with earlier estimates by O. K. Andersen[22] from a bandstructure analysis. An experimental and theoretical comparison between $T_2$ and $T_{2\perp}$ might help to identify the decoupling transition.

## 4. ACKNOWLEDGMENTS

We would like to thank T. M. Rice for many useful discussions inspiring this work. In addition we would like to thank the Zürich NMR group for the enjoyable discussions. A. W. S. acknowledges support from the DOE grant DE-FG03-85ER45197.

## REFERENCES

1. For a review see the articles of D. Pines and D. J. Scalapino in the proceedings of the M$^2$SHTC meeting in Grenoble, to be published in Physica **C**
2. S. Chakravarty, B. I. Halperin, and D. R. Nelson, Phys. Rev. Lett. **60**, 1057 (1988) and S. Chakravarty, B. I. Halperin, and D. R. Nelson, Phys. Rev. **B 39**, 7443 (1988).
3. A. J. Millis and H. Monien, Phys. Rev. Lett. **70**, 2810 (1993).
4. A. Sokol and D. Pines, Phys. Rev. Lett. **71**, 2813 (1993).
5. S. Sachdev and J. Ye, Phys. Rev. Lett. **69**, 2411 (1992).
6. A. V. Chubukov, S. Sachdev, and A. Sokol, Phys. Rev. **B 49**, 9052 (1994) and A. V. Chubukov, S. Sachdev, and J. Ye, Phys. Rev. **B 49**, 11919 (1994).
7. A. Sokol, R. L. Glenister, and R. R. P. Singh, Phys. Rev. Lett. **72**, 1549 (1994).
8. A. W. Sandvik and D. J. Scalapino, Phys. Rev. Lett. **72**, 2777 (1994).
9. J. M. Tranquada *et al.*, Phys. Rev. **B 46**, 5561 (1992).
10. K. Hida, J. Phys. Soc. Jpn. **61**, 1013 (1992).
11. L. B. Ioffe, A. I. Larkin, A. J. Millis and B. L. Altshuler, JETP Lett. **59**, 65 (1994)
12. M. U. Ubbens and P. A. Lee, Phys. Rev. **49**, 6853 (1994) and M. U. Ubbens and P. A. Lee, Phys. Rev. **B50**, 438 (1994).
13. M. J. Lercher and J. M. Wheatley, Phys. Rev. **B49**, 736 (1994)
14. H. Monien and T. M. Rice, cond-mat 9406118, in the proceedings of the M$^2$SHTC meeting in Grenoble, to be published in Physica **C**
15. R. Stern, M. Mali, I. Mangelschots, J. Roos, and D. Brinkmann, J.-Y. Genoud, T. Graf, and J. Mueller, Phys. Rev. **B50**, 426 (1994)
16. C. H. Pennington and C. P. Slichter, Phys. Rev. Lett. **66**, 381 (1991).
17. A. J. Millis, H. Monien, and D. Pines, Phys. Rev. **B 42**, 996 (1991).
18. K. Hida, J. Phys. Soc. Jpn. **59**, 2230 (1990).
19. A. J. Millis and H. Monien, Physical Review **B**, in print.
20. A. W. Sandvik, J. Phys. **A25**, 3667 (1992).
21. R. Stern and M. Mali, unpublished
22. O. K. Andersen, private communication.